\documentclass[twocolumn,prl,amsmath,amssymb,showpacs,superscriptaddress]{revtex4}
\usepackage{graphicx}
\usepackage{dcolumn}
\usepackage{bm}
\usepackage{latexsym}
\usepackage{amstext}
\usepackage{amsxtra}

\begin{document}
\title{$d-$wave collapse and explosion of a dipolar Bose-Einstein condensate}
\author{T.~Lahaye}
\affiliation{5. Physikalisches Institut, Universit\"at Stuttgart, Pfaffenwaldring 57, 70550 Stuttgart, Germany}
\author{J.~Metz}
\affiliation{5. Physikalisches Institut, Universit\"at Stuttgart, Pfaffenwaldring 57, 70550 Stuttgart, Germany}
\author{B.~Fr\"ohlich}
\affiliation{5. Physikalisches Institut, Universit\"at Stuttgart, Pfaffenwaldring 57, 70550 Stuttgart, Germany}
\author{T.~Koch}
\affiliation{5. Physikalisches Institut, Universit\"at Stuttgart, Pfaffenwaldring 57, 70550 Stuttgart, Germany}
\author{M.~Meister}
\affiliation{5. Physikalisches Institut, Universit\"at Stuttgart, Pfaffenwaldring 57, 70550 Stuttgart, Germany}
\author{A.~Griesmaier}
\affiliation{5. Physikalisches Institut, Universit\"at Stuttgart, Pfaffenwaldring 57, 70550 Stuttgart, Germany}
\author{T.~Pfau}
\affiliation{5. Physikalisches Institut, Universit\"at Stuttgart, Pfaffenwaldring 57, 70550 Stuttgart, Germany}
\author{H.~Saito}
\affiliation{Department of Applied Physics and Chemistry, The University of Electro-Communications, Tokyo 182-8585, Japan}
\author{Y.~Kawaguchi}
\affiliation{Department of Physics, University of Tokyo, Tokyo 113-0033, Japan}
\author{M.~Ueda}
\affiliation{Department of Physics, University of Tokyo, Tokyo 113-0033, Japan}
\affiliation{ERATO Macroscopic Quantum Project, JST, Tokyo 113-8656, Japan}
\date{\today}

\begin{abstract}
We investigate the collapse dynamics of a dipolar condensate of $^{52}$Cr atoms when the $s-$wave scattering length characterizing the contact interaction is reduced below a critical value. A complex dynamics, involving an anisotropic, $d-$wave symmetric explosion of the condensate, is observed. The atom number decreases abruptly during the collapse. We find good agreement between our experimental results and a those of a numerical simulation of the three-dimensional Gross-Pitaevskii equation, including contact and dipolar interactions as well as three-body losses. The simulation indicates that the collapse induces the formation of two vortex rings with opposite circulations.
\end{abstract}

\pacs{03.75.Kk,03.75.Lm}

\maketitle

The underlying symmetries of physical systems often determine the nature and dynamics of macroscopic quantum states. For example, the difference between isotropic and $d-$wave pairing of electrons, in conventional and high-$T_{\rm c}$ superconductors, respectively, leads to fundamentally different properties~\cite{ref1}. Degenerate quantum gases are usually dominated by isotropic ($s-$wave) contact interactions. \emph{Dipolar} quantum gases (i.e. in which the dipole-dipole interaction (DDI) between permanent dipole moments play a significant or even dominant role) are governed by the $d-$wave symmetry of the long-range DDI, which gives rise to novel properties.

Examples of fascinating predictions for polarized dipolar quantum gases range from a roton-maxon spectrum~\cite{ref3} for the elementary excitations of a quasi two-dimensional dipolar Bose-Einstein condensate (BEC), to the existence of novel quantum phases (such as a `checkerboard' insulator, or a supersolid) for dipolar quantum gases in optical lattices~\cite{qphases}. The DDI also modifies the hydrodynamic equations describing the dynamics of a BEC, which has been probed experimentally by studying the expansion of the cloud when released from the trap~\cite{hd}. Unusual, structured shapes for the BEC have been predicted~\cite{ref12,ref13}. In the unpolarized case, the DDI dramatically enriches the physics of spinor BECs~\cite{spinor}. Ultracold dipolar fermions also have fascinating properties~\cite{ddi-fermions}.

A striking example of the new properties of dipolar BECs is given by their stability, which, contrary to the case of contact interaction, depends strongly on the trap geometry. Consider a pancake-shaped trap with the dipole moments of the particles oriented perpendicular to the plane of the trap. The DDI is then essentially repulsive, and the BEC is stable, independently of the atom number. In contrast, a cigar-shaped trap cannot stabilize a purely dipolar BEC. We experimentally studied~\cite{ref2} this geometry-dependent stability of a dipolar quantum gas by using a $^{52}$Cr BEC, and mapped out the stability diagram of the condensate as a function of the scattering length $a$ (characterizing the contact interaction) and the trap aspect ratio. In the case of a pure contact interaction, crossing the stability border into the unstable regime $a<0$ leads to a collapse of the BEC~\cite{ref18,ref19,ref20}. This gives rise to an interesting dynamics involving a fast implosion of the condensate followed by the formation of energetic `bursts' of atoms~\cite{ref21}, or the formation of soliton trains~\cite{ref22,ref23}.

In this Letter, we investigate experimentally the collapse dynamics of a dipolar $^{52}$Cr BEC when the scattering length $a$ is decreased (by means of a Feshbach resonance) below the critical value $a_{\rm crit}$ for stability~\cite{ref2}. We observe a rich dynamics on a timescale shorter than the trap period, with the formation of an expanding structure featuring a $d-$wave symmetry. We study the atom number in the condensate as a function of time, and find an abrupt decrease due to inelastic losses. Finally, we compare our experimental results with a three-dimensional numerical simulation of the Gross-Pitaevskii equation (GPE) including both contact and dipolar interactions, as well as three-body losses. Such a generalized GPE with three-body losses has been demonstrated to explain the main features of `Bose-nova' experiments with $^{85}$Rb~\cite{ueda2003}. Here we generalize this model to include the DDI. As shown later, this generalized model accounts very well for the observed $d-$wave collapse.

The experimental setup to produce a $^{52}$Cr BEC above the Feshbach resonance located at a magnetic field $B_0 \simeq 589$~G has been described elsewhere~\cite{ref2}. Close to the resonance, the scattering length $a$ varies with the applied magnetic field $B$ as
$$
a(B)=a_{\rm bg}\left(1-\frac{\Delta}{B-B_0}\right),
$$
where $\Delta\simeq1.5$~G is the resonance width, and $a_{\rm bg}\simeq 100 \,a_0$ the background scattering length ($a_0$ is the Bohr radius). We calibrate the variation $a(B)$ of the scattering length by measuring, after expansion, the BEC size and atom number~\cite{ref2}. The reduction of $a$ close to $B_0 + \Delta$ is accompanied by inelastic losses. By measuring the $1/e$ lifetime and the density of the BEC close to resonance, we estimate the three-body loss coefficient to be constant for the range of scattering lengths ($5\leqslant a/a_0\leqslant30$) studied here, with a value $L_3 \sim 2\times 10^{-40} \;{\rm m}^6/{\rm s}$.

\begin{figure}[t]
\centerline{\includegraphics[width=80mm]{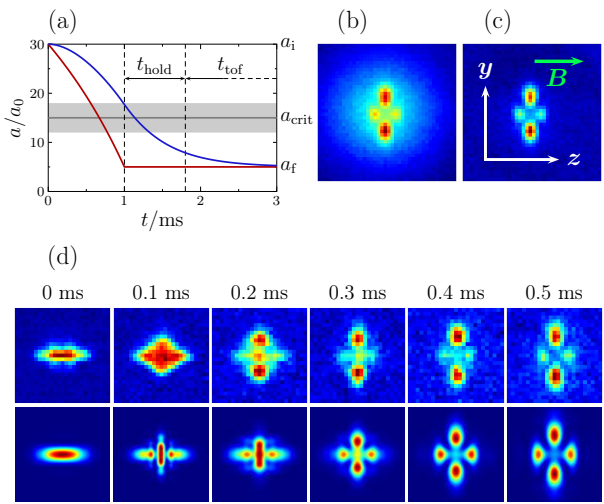}} \caption{(color). Collapse dynamics of the dipolar condensate. (a) Timing of the experiment. The red curve represents the time variation of the scattering length $a(t)$ one would have in the absence of eddy currents, while the blue curve is obtained by taking them into account (see text). (b) Sample absorption image of the collapsed condensate for $t_{\rm hold} = 0.4$~ms, after 8~ms of time of flight, showing a `cloverleaf' pattern on top of a broad thermal cloud. This image was obtained by averaging 60 pictures taken under the same conditions. (c) Same image as (b) with the thermal cloud subtracted. In (b) and (c) the field of view is 270 $\mu$m by 270 $\mu$m. The green arrow indicates the direction of the magnetic field. (d) Series of images of the condensate for different values of $t_{\rm hold}$ (upper row) and results of the numerical simulation without adjustable parameters (lower row); the field of view is 130 $\mu$m by 130 $\mu$m.}
\label{fig1}
\end{figure}

To study the collapse dynamics, we first create a BEC of typically 20,000 atoms in a trap with frequencies $(\nu_x,\nu_y,\nu_z) \simeq (660,400,530)$~Hz at a magnetic field $\sim 10$~G above the Feshbach resonance, where the scattering length is $a \simeq 0.9 \,a_{\rm bg}$. We then decrease $a$ by ramping down $B$ linearly over 8~ms to a value $a_{\rm i} = 30 \,a_0$ which still lies well above the critical value for collapse, measured to be at $a_{\rm crit}\simeq (15\pm3) \,a_0$ [shaded area on Fig. 1(a)] for our parameters~\cite{ref2}. This ramp is slow enough to be adiabatic ($\dot{a}/a\ll \nu_{x,y,z}$), so that the BEC is not excited during it. After 1~ms waiting time, $a$ is finally ramped down rapidly to $a_{\rm f}  = 5 \,a_0$, which is below the collapse threshold. For this, we ramp linearly in 1~ms the current $I(t)$ in the coils providing the magnetic field $B=\alpha I$. However, due to eddy currents in the metallic vacuum chamber, the actual value of $B(t)$ and hence that of $a(t)$ change in time as depicted in blue on Fig. 1(a). To obtain this curve, we used Zeeman spectroscopy to measure the step response of $B(t)$ to a jump in the current $I(t)$ (corresponding to a $\sim15$~G change in $B$), and found that the resulting $B(t)$ is well described if $\tau\dot{B}+B=\alpha I(t)$ holds, with $\tau\simeq 0.5$~ms. From this equation and the measured $I(t)$ we determine the actual $a(t)$.

After the ramp, we let the system evolve for an adjustable time $t_{\rm hold}$ and then the trap is switched off. Note that the origin of $t_{\rm hold}$ corresponds to the end of the ramp in $I(t)$. Because of eddy currents, $t_{\rm hold}=0$ about 0.2~ms \emph{before} the time at which the scattering length crosses $a_{\rm crit}$. However, as we shall see below, even for $t_{\rm hold}<0.2$~ms a collapse (happening not in trap, but during the time of flight) is observed, since during expansion the scattering length continues to evolve towards $a_{\rm f}$. The large magnetic field along $z$ is rapidly turned off (in less than 300~$\mu$s) after 4~ms of expansion, and the condensate expands for another 4~ms in an 11~G field pointing in the $x$ direction, before being imaged by absorption of a resonant laser beam propagating along $x$. Changing the direction of the field allows us to use the maximum absorption cross-section for the imaging (if the latter was done in high field, the absorption cross-section would be smaller, thus reducing the signal to noise ratio of the images). We checked that this fast switching has no influence on the condensate dynamics. We observe that the atomic cloud has a clear bimodal structure, with a broad isotropic thermal cloud, well fitted by a Gaussian, and a much narrower, highly anisotropic central feature, interpreted as the remnant BEC [see Fig. 1(b) and (c)].

\begin{figure}[t]
\centerline{\includegraphics[width=80mm]{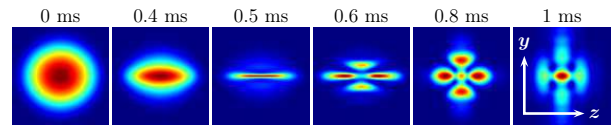}} \caption{(color). In-trap column density obtained in the simulation, for different $t_{\rm hold}$. The field of view is 5 $\mu$m by 5 $\mu$m. Due to the DDI, the condensate collapses radially, acquiring the shape of a very thin cigar elongated along $z$. At $t_{\rm hold}\simeq0.5$~ms, the collapse occurs, and immediately after, the cloud starts to expand radially.}
\label{fig2}
\end{figure}

The upper row of Fig. 1(d) shows the time evolution of the condensate when varying $t_{\rm hold}$. The images were obtained by averaging typically five absorption images taken under the same conditions; the thermal background was subtracted, and the color scale was adjusted separately for each $t_{\rm hold}$ for a better contrast. From an initial shape elongated along the magnetization direction $z$, the condensate rapidly develops a complicated structure with an expanding, torus-shaped part close to the $z = 0$ plane. Interestingly, the angular symmetry of the condensate at some specific times (e.g. at $t_{\rm hold} = 0.5$~ms) is reminiscent of the $d-$wave angular symmetry $1-3\cos^2\theta$ of the DDI. For larger values of $t_{\rm hold}$, we observe that the condensate `refocuses' due to the presence of the trap~\cite{epaps}.

The lower row of Fig. 1(d) shows the column density $\int \left|\psi(\bm{r})\right|^2\,{\rm d}x$ (where $\psi(\bm{r})$ is the order parameter of the condensate after time of flight) obtained from a numerical simulation of the three-dimensional GPE
\begin{eqnarray*}
i\hbar\frac{\partial \psi}{\partial t} &=&\left[\frac{-\hbar^2}{2m}\triangle +V_{\rm trap}+\int U(\bm{r}-\bm{r}',t)\left|\psi(\bm{r}',t)\right|^2\,{\rm d}\bm{r}'\right.\\&-&\left.\frac{i\hbar L_3}{2}\left|\psi\right|^4\right]\psi,
\end{eqnarray*}
where
$$
U(\bm{r},t)=\frac{4\pi\hbar^2a(t)}{m}\delta(\bm{r})+\frac{\mu_0\mu^2}{4\pi}\frac{1-3\cos^2\theta}{r^3}
$$
stands for the contact and dipolar interactions, $\theta$ being the angle between $\bm{r}$ and the direction of polarization. Here $m$ is the atomic mass, $\mu_0$ the permeability of vacuum, and $ \mu = 6 \mu_{\rm  B}$ the magnetic moment of a Cr atom ($\mu_{\rm B}$ is the Bohr magneton). The non-unitary term proportional to $L_3$ describes three-body losses. The scattering length $a(t)$ is changed according to the blue curve in Fig. 1(a) and the trap potential $V_{\rm trap}$ is switched off at the beginning of the 8~ms time of flight. For the simulation, space is discretized into a $128\times128\times128$ mesh with a step size of 70~nm. For the kinetic part, the Crank-Nicolson scheme is used for the time evolution to avoid  numerical instability. For the interaction part, the convolution integral is calculated using a fast Fourier transform. After the trap is switched off, the mesh is extended to $512\times512\times512$ to describe the expansion of the cloud. When the density becomes low enough so that the nonlinear terms of the GPE can be neglected, the free expansion propagator is used to give the final time-of-flight images.  The agreement between the experimental data and the simulation, performed without any adjustable parameter, is excellent.

The cloverleaf patterns seen in Fig. 1(b)-(d) are caused by the anisotropic collapse and the subsequent dynamics of the system. Figure~2 shows the in-trap evolution of the condensate as a function of $t_{\rm hold}$. The mechanism of the condensate `explosion' is as follows~\cite{ref24}: When the atomic density grows due to the attractive interaction, three-body losses predominantly occur in the high-density region. The centripetal force is then decreased, and the atoms that gathered in this narrow central region are ejected due to the `quantum pressure' arising from the uncertainty principle. The kinetic energy is supplied by the loss of the negative interaction energy. For the contact interaction, the collapse and subsequent atomic `explosion' is isotropic~\cite{ref24}. In the present case, the collapse occurs mainly in the $x-y$ direction due to anisotropy of the DDI (in the absence of inelastic losses, the condensate would indeed become an infinitely thin cigar-shaped cloud along $z$, see Fig. 3 of Ref.\cite{ref2}, and the in-trap image at $t_{\rm hold}=0.5$~ms in Fig.~2), and therefore the condensate `explodes' essentially radially, producing the anisotropic shape of the cloud. The numerical simulation reveals that, for $t_{\rm hold} < 0.5$~ms, the collapse observed in Fig. 1(d) occurs not during the holding time but during the time-of-flight. We stress that in the absence of three-body losses, the explosion following the collapse would not be observed.

\begin{figure}[t]
\centerline{\includegraphics[width=70mm]{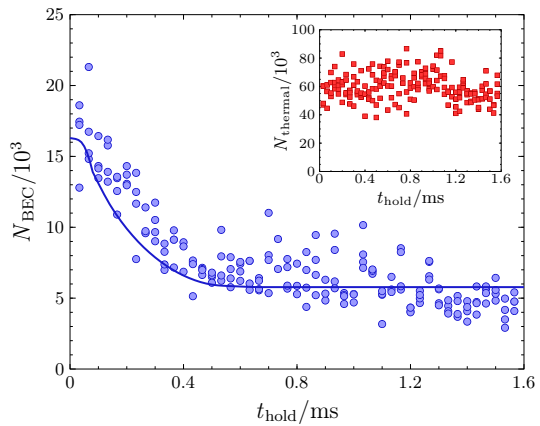}} \caption{(color). Atom losses during collapse. Blue circles: atom number $N_{\rm BEC}$ in the condensate as a function of $t_{\rm hold}$. The solid curve is the  result of the simulation for $L_3 = 2 \times 10^{-40} \;\rm{m}^6/\rm{s}$, without any adjustable parameter. Inset: the atom number $N_{\rm thermal}$ in the thermal cloud (red squares) remains essentially constant during the collapse.}
\label{fig3}
\end{figure}

From the images, the atom number $N_{\rm BEC}$ in the condensate is obtained by integrating the optical density. Blue circles in Fig. 3 show $N_{\rm BEC}$ as a function of $t_{\rm hold}$. The BEC atom number is initially $N_{\rm BEC}(0) \simeq 16,000$ and decreases toward its asymptotic value $\sim 6,000$. Over the same time scale, the atom number $N_{\rm thermal}$ in the thermal cloud (inset of Fig. 3) stays constant. The size of the thermal cloud after expansion is also constant over this period. This suggests that the thermal cloud does not play any significant role in the collapse dynamics. For $t_{\rm hold} < 0.5$~ms, the collapse actually occurs during the time of flight, which explains the gradual decay of $N_{\rm BEC}(t_{\rm hold})$, and why the atom losses are not as large as those when the collapse occurs in trap (in the latter case, 70\% of the atoms are lost). The missing atoms have very likely escaped from the trap as energetic molecules and atoms produced in three-body collisions. This is confirmed by the fact that the simulation gives a $N_{\rm BEC}$($t_{\rm hold}$) curve (solid line in Fig. 3) which matches well the experimental data. Experimental uncertainties in the parameters used in the simulation (trap frequencies, values of $L_3$ and $\tau$) probably explain the small discrepancy between the experiment and the numerical results.

The numerical simulation gives access not only to the density $|\psi(\bm{r})|^2$, but also to the phase $S(\bm{r})$ of the order parameter $\psi$ (i.e. to the velocity field $\bm{v}=\hbar\bm{\nabla}S/m$) and reveals the generation of vortex rings~\cite{ref25,ref26} of charge $\pm1$. Figure 4(a) shows an in-trap iso-density surface of a condensate at $t_{\rm hold} = 0.8$~ms and the location of the vortex rings (shown as red curves). Comparing Fig. 4(a) and Fig. 2 (at $t_{\rm hold} = 0.8 $~ms), we find that the topological defects encircle the two `leaves' of the `clover'.  Figure 4(b) shows the velocity field $\bm{v}(\bm{r})$ in the $x = 0$ plane. The atoms ejected in the $x-y$ plane flow outward, while the atoms near the $z$ axis still flow inward, giving rise to the circulation. Thus, the vortex-ring formation is specific to the $d-$wave collapse induced by the DDI. Although the vortex rings are not observed directly in the experiment, the excellent agreement between the experiment and simulation in Figs. 1 and 3 strongly suggests the creation of vortex rings during the collapse of the condensate.

\begin{figure}[t]
\centerline{\includegraphics[width=75mm]{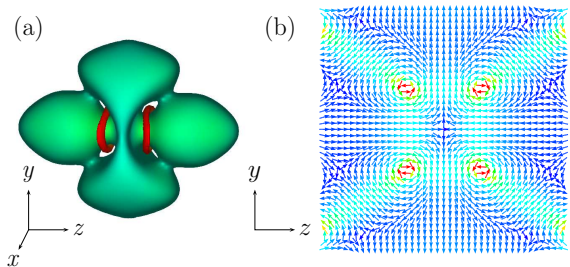}} \caption{(color). Vortex rings predicted by the numerical simulation. (a) Iso-density surface of an in-trap condensate at $t_{\rm hold} = 0.8$~ms. The topological defects are shown by the red rings. (b) Velocity field of the atomic flow in the $x = 0$ plane at $t_{\rm hold} = 0.8$~ms. The field of view is 2.5 $\mu$m by 2.5 $\mu$m; the color scale represents the velocity (red is faster).}
\label{fig4}
\end{figure}

In conclusion, we have investigated the collapse dynamics of a dipolar BEC. Contrary to the case of an isotropic contact interaction, the DDI induces the formation of a structured cloud featuring a $d-$wave symmetry. The collapse dynamics is quantitatively reproduced by numerical simulations of the GPE without any adjustable parameter. An interesting subject for future studies is the dependence of the collapse dynamics on the trap geometry: one may wonder if the condensate would collapse in the same way if initially trapped in a very anisotropic (e.g., pancake-shaped) trap. A natural extension of this work would involve detecting, e.g. by interferometric methods~\cite{ref27,ref28}, the vortex rings predicted by the simulation. Finally, whether one can nucleate stable vortex rings by initiating the collapse, and then changing $a$ back to a value corresponding to a stable BEC, is a question which certainly deserves further investigations.

We thank H. P. B\"uchler, S. Giovanazzi, L. Santos and G. V. Shlyapnikov for useful discussions. We acknowledge support by the German Science Foundation (SFB/TRR 21 and SPP 1116) and the EU (Marie-Curie Grant MEIF-CT-2006-038959 to T. L.). H. S., Y. K. and M. U. acknowledge support by the Ministry of Education, Culture, Sports, Science and Technology of Japan (Grants-in-Aid for Scientific Research No. 17071005 and No. 20540388, the 21st century COE program on `Nanometer-Scale Quantum Physics') and by the Matsuo Foundation.

\end{document}